\documentclass[runningheads]{llncs}

\usepackage{graphicx}
\usepackage{array}
\usepackage{hyperref}

\usepackage{float}
\begin{document}
\title{nnDetection: A Self-configuring Method for Medical Object Detection}
\author{Michael Baumgartner$^{1\star}$, Paul F. Jäger$^{2\star}$, Fabian Isensee$^{1,3}$, Klaus H. Maier-Hein$^{1,4}$}

\authorrunning{Baumgartner and J{\"a}ger et al.}

\renewcommand{\thefootnote}{\fnsymbol{footnote}}
\footnotetext[1]{Equal contribution.}


\institute{%
$^1$Division of Medical Image Computing, German Cancer Research Center, Heidelberg, Germany\\
$^2$Interactive Machine Learning Group, German Cancer Research Center\\
$^3$HIP Applied Computer Vision Lab, German Cancer Research Center\\
$^4$Pattern Analysis and Learning Group, Heidelberg University Hospital\\
\email{m.baumgartner@dkfz.de}}

\maketitle              

\begin{abstract}
Simultaneous localisation and categorization of objects in medical images, also referred to as medical object detection, is of high clinical relevance because diagnostic decisions often depend on rating of objects rather than e.g. pixels. For this task, the cumbersome and iterative process of method configuration constitutes a major research bottleneck. 
Recently, nnU-Net has tackled this challenge for the task of image segmentation with great success. Following nnU-Net’s agenda, in this work we systematize and automate the configuration process for medical object detection. The resulting self-configuring method, nnDetection, adapts itself without any manual intervention to arbitrary medical detection problems while achieving results en par with or superior to the state-of-the-art. We demonstrate the effectiveness of nnDetection on two public benchmarks, ADAM and LUNA16, and propose 11 further medical object detection tasks on public data sets for comprehensive method evaluation. Code is at  \url{https://github.com/MIC-DKFZ/nnDetection}.
\end{abstract}

\section{Introduction}
Image-based diagnostic decision-making is often based on rating objects and rarely on rating individual pixels. This process is well reflected in the task of medical object detection, where entire objects are localised and rated. Nevertheless, semantic segmentation, i.e. the categorization of individual pixels, remains the predominant approach in medical image analysis with 70$\%$ of biomedical challenges revolving around segmentation \cite{maier_hein_18}. To be of diagnostic relevance, however, in many use-cases segmentation methods require ad-hoc postprocessing that aggregates pixel predictions to object scores. This can negatively affect performance compared to object detection methods that already solve these steps within their learning procedure \cite{jaeger_20}.

\noindent Compared to a basic segmentation architecture like the U-Net, the set of hyper-parameters in a typical object detection architecture is extended by an additional detection head with multiple loss functions including smart sampling strategies ("hard negative mining"), definition of size, density and location of prior boxes ("anchors"), or the consolidation of overlapping box predictions at test time ("weighted box clustering"). This added complexity might be an important reason for segmentation methods being favoured in many use-cases. It further aggravates the already cumbersome and iterative process of method configuration, which currently requires expert knowledge, extensive compute resources, sufficient validation data, and needs to be repeated on every new tasks due to varying data set properties in the medical domain \cite{isensee_21}.  

\noindent Recently, nnU-net achieved automation of method configuration for the task of biomedical image segmentation by employing a set of fixed, rule-based, and empirical parameters to enable fast, data-efficient, and holistic adaptation to new data sets \cite{isensee_21}. In this work, we follow the recipe of nnU-Net to systematize and automate method configuration for medical object detection. Specifically, we identified a novel set of fixed, rule-based, and empirical design choices on a diverse development pool comprising 10 data sets. We further follow nnU-Net in deploying a clean and simple base-architecture: Retina U-Net \cite{jaeger_20}. The resulting method, which we call nnDetection, can now be fully automatically deployed on arbitrary medical detection problems without requiring compute resources beyond standard network training. 

\noindent Without manual intervention, nnDetection sets a new state of the art on the nodule-candidate-detection task of the well-known LUNA16 benchmark and achieves competitive results on the ADAM leaderboard. To address the current lack of public data sets compared to e.g. medical segmentation, we propose a new large-scale benchmark totaling 13 data sets enabling sufficiently diverse evaluation of medical object detection methods. To this end, we identified object detection tasks in data sets of existing segmentation challenges and compare nnDetection against nnU-Net (with additional postprocessing for object scoring) as a standardized baseline.

\noindent With the hope to foster increasing research interest in medical object detection, we make nnDetection publicly available (including pre-trained models and object annotations for all newly proposed benchmarks) as an out-of-the-box method for state-of-the-art object detection on medical images, a framework for novel methodological work, as well as a standardized baseline to compare against without manual effort.
\section{Methods}
\begin{figure}[!htbp]
    \centering
    \includegraphics[width=\linewidth]{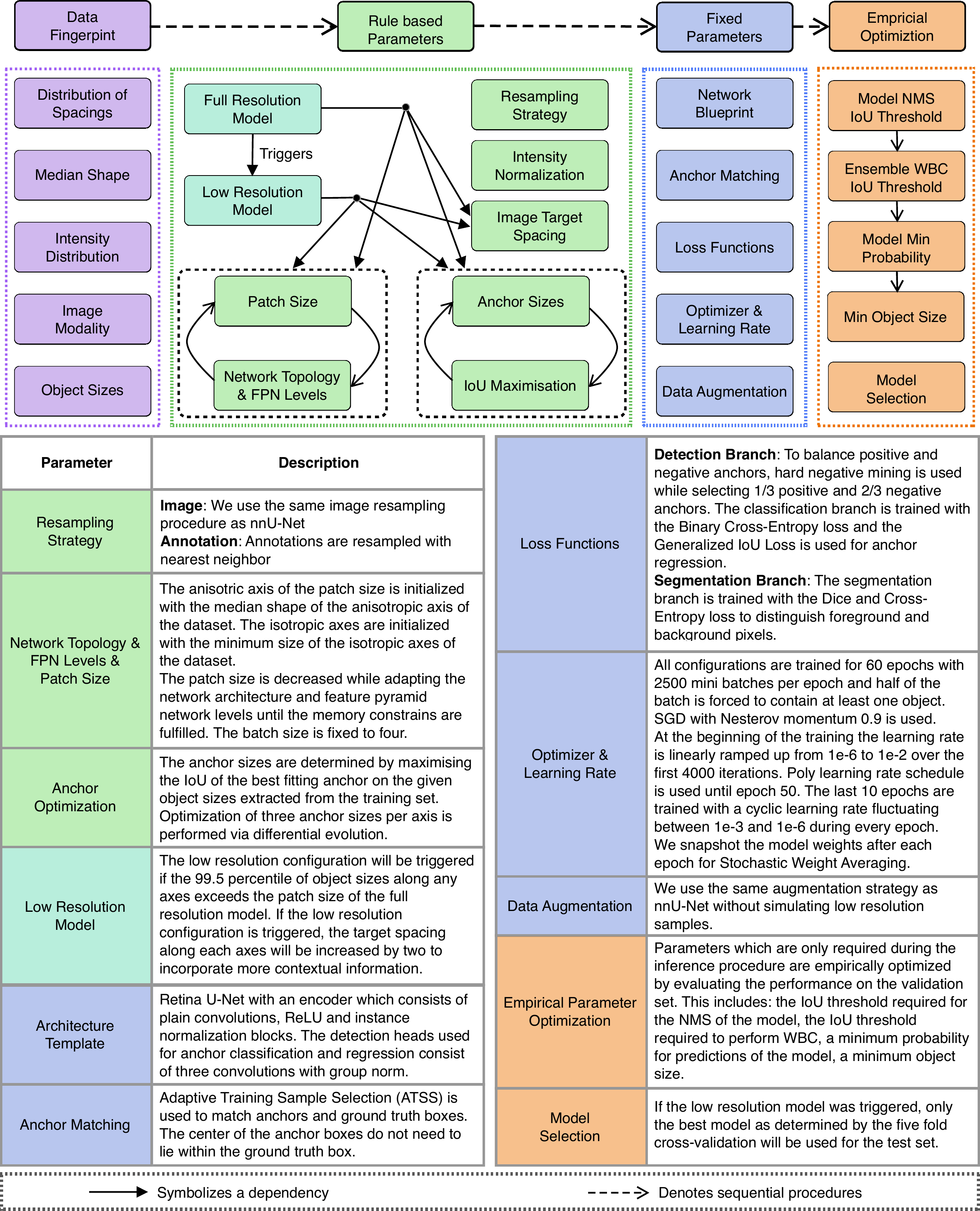}
    \caption{
    Overview of the high level design choices and mechanisms of nnDetection (For details and reasonings about all design decisions we refer to our code repository at \url{https://github.com/MIC-DKFZ/nnDetection}). Due to the high number of dependencies between parameters, only the most important ones are visualized as arrows. Given a new medical object detection task, a fingerprint covering relevant data set properties is extracted (purple). Based on this information, a set of heuristic rules is executed to determine the rule-base parameters of the pipeline (green). These rules act in tandem with a set of fixed parameters which do not require adaptation between data sets (blue). After training, empirical parameters are optimized on the training set (orange). All design choices were developed and extensively evaluated upfront on our development pool, thus ensuring robustness and enabling rapid application of nnDetection to new data sets without requiring extensive additional compute resources.}
    \label{fig:nnDetection}
\end{figure}
Fig.~\ref{fig:nnDetection} shows how nnDetection systematically addresses the configuration of entire object detection pipelines and provides a comprehensive list of design choices.

\noindent\textbf{nnDetection development.} To achieve automated method configuration in medical object detection, we roughly follow the recipe outlined in nnU-Net, where domain knowledge is distilled in the form of fixed, rule-based, and empirical parameters. Development was performed on a pool of 10 data sets (see supplementary material).

\noindent\textit{Fixed Parameters:} (For a comprehensive list see Fig.~\ref{fig:nnDetection}). First, we identified design choices that do not require adaptation between data sets and optimized a joint configuration for robust generalization on our 10 development data sets. We opt for Retina U-Net as our architecture template, which builds on the simple RetinaNet to enable leveraging of pixel-level annotations \cite{jaeger_20}, and leave the exact topology (e.g. kernel sizes, pooling strides, number of pooling operations) to be adapted via rule-based parameters. To account for varying network configurations and object sizes across data sets we employ adaptive training sample selection \cite{zhang_20} for anchor matching. However, we discarded the requirement as to which the center point of selected anchors needs to lie inside the ground truth box because, as we found it often resulted in the removal of all positive anchors for small objects. Furthermore, we increased the number of anchors per position from one to 27, which we found improves results especially on data sets with few objects.

\noindent\textit{Rule-based Parameters:} (For a comprehensive list see Fig.~\ref{fig:nnDetection}). Second, for as many of the remaining decisions as possible, we formulate explicit dependencies between the Data Fingerprint and design choices in the form of interdependent heuristic rules. Compared to nnU-Net our Data Fingerprint additionally extracts information about object sizes (see Figure~\ref{fig:nnDetection}). We use the same iterative optimization process as nnU-Net to determine network topology parameters such as kernel sizes, pooling strides, and the number of pooling operations, but fixed the batch size at 4 as we found this to improve training stability. Similar to nnU-Net, an additional low-resolution model is triggered to account for otherwise missing context in data sets with very large objects or high resolution images. Finding an appropriate anchor configuration is one of the most important design choices in medical object detection \cite{zlocha_19,redmon_17}. Following Zlocha et al. \cite{zlocha_19}, we iteratively maximize the intersection over union (IoU) between anchors and ground-truth boxes. In contrast to their approach, we found performing this optimization on the training split instead of the validation split led to more robust anchor configurations due to a higher number of samples. Also, we fit three anchor sizes per axis and use the euclidean product to produce the final set of anchors for the highest resolution pyramid level the detection head operates on.

\noindent\textit{Empirical Parameters:} (For a comprehensive list see Fig.~\ref{fig:nnDetection}). Postprocessing in object detection models mostly deals with clustering overlapping bounding box predictions. There are different sources for why predictions might overlap. The inherent overlap of predictions from dense anchors is typically accounted for by Non-maximum Suppression (NMS).
Due to limited GPU memory, nnDetection uses sliding window inference with overlapping patches. Overlaps across neighboring patches are clustered via NMS while weighing predictions near the center of a patch higher than predictions at the border. To cluster predictions from multiple models or different test time augmentations Weighted Box Clustering \cite{jaeger_20} is used. Empirical Parameters which are only used at test time (see a full list in the Table in Figure~\ref{fig:nnDetection}) are optimized empirically on the validation set. Due to their interdependencies, nnDetection uses a pre-defined initialization of the parameters and sequentially optimizes them by following the order outlined in Fig.~\ref{fig:nnDetection}. If the low resolution model has been triggered, the best model will be selected empirically for testing based on the validation results. 

\noindent\textbf{nnDetection application.} Given a new data set, nnDetection runs automatic configuration without manual intervention. Thus, no additional computational cost beyond a standard network training procedure is required apart from the few required empirical choices. First, nnDetection extracts the Data Fingerprint and executes the heuristic rules to determine the rule-based parameters. Subsequently, the full-resolution and, if triggered, the low-resolution model will be trained via five-fold cross-validation. After training, empirical parameters are determined and the final prediction is composed by ensembling the predictions of the five models obtained from cross-validation of the empirically selected configuration. We evaluate the generalization ability of nnDetection’s automated configuration on 3 additional data sets (see supplementary material).

\noindent\textbf{nnU-Net as an object detection baseline.} 
\label{sec:nnunet}
Our first nnU-Net baseline, called \textit{nnU-Net Basic}, reflects the common approach to aggregating pixel predictions: Argmax is applied over softmax predictions, followed by connected component analysis per foreground class, and finally an object score per component is obtained as the maximum pixel softmax score of the assigned category. \textit{nnU-Net Plus:} To ensure the fairest possible comparison, we enhance the baseline by empirically choosing the following postprocessing parameters based on the training data for each individual task: Replacement of argmax by a  minimum threshold on the softmax scores to be assigned to a component, a threshold on the minimum number of pixels per object, and the choice of the aggregation method (max, mean, median, $95\%$ percentile). During our experiments on LIDC \cite{armato_11} we observed convergence issues of nnU-Net. Thus, we identified an issue with numerical constants inside the Dice loss and were able to improve results significantly by removing those.

\begin{figure}[t]
    \centering
    \includegraphics[width=\textwidth]{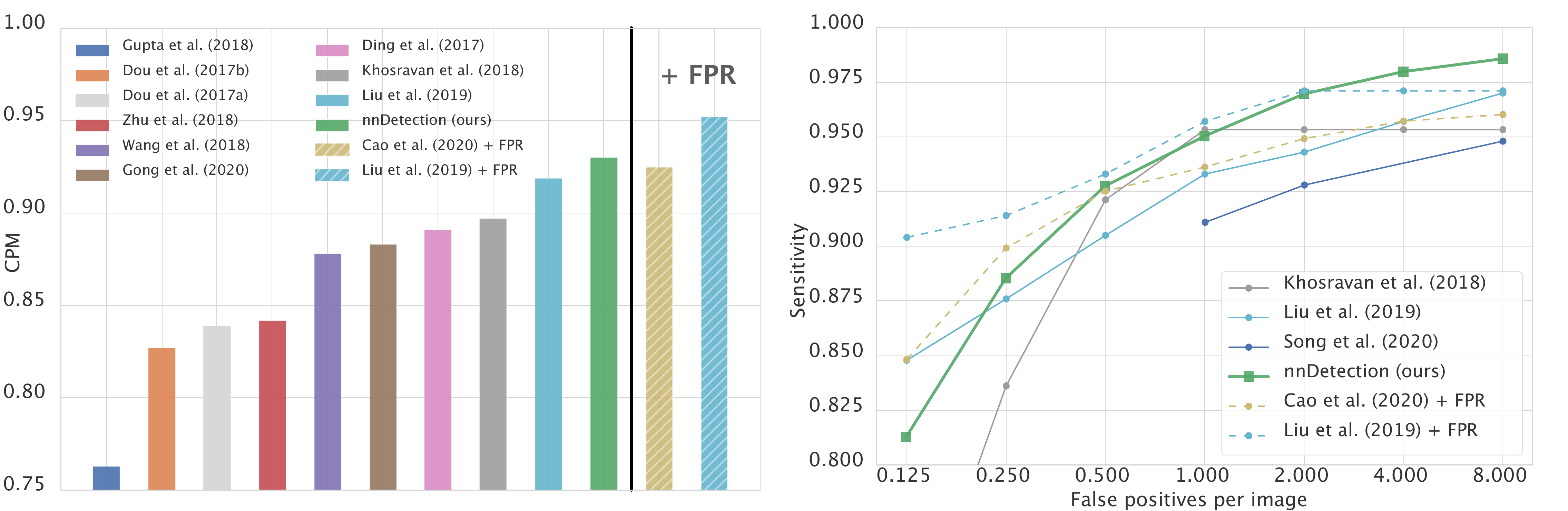}
    \caption{\textit{Left:} nnDetection outperforms all competing  approaches on the nodule-candidate-detection task of LUNA16 and is only beaten by Liu et al. \cite{liu_19} in the general task, where additional False Positive Reduction (FPR) models are employed (we consider such task-specific intervention to be out of scope for this work). \textit{Right:} FROC curves for the top 7 methods. Starting from 1/4 False Positives per Scan, nnDetection outperforms Liu et al. \cite{liu_19} without FPR.}
    \label{fig:results_luna}
\end{figure}

\section{Experiments and Results}
\textbf{Proposed benchmark for medical object detection.} Recently, strong evidence has been provided for the importance of evaluating segmentation methods on a large and diverse data set pool \cite{isensee_21}. This requirement arises from volatility of evaluation metrics caused by limited data set size as well as considerable label noise in the medical domain. Furthermore, covering data set diversity prevents general methodological claims from being overfit to specific tasks. We argue these aspects directly translate to medical object detection and thus propose a new benchmark based a diverse pool of 13 existing data sets. Since public benchmarks are less abundant compared to segmentation tasks, we identified object detection tasks in 5 data sets of existing segmentation challenges (where we focus on detecting tumors and consider organs as background, see supplementary material for details). To generate object annotations from pixel-level label maps, we performed connected component analysis and discarded all objects with a diameter less than $3mm$. Further, object annotations originating from obvious segmentation errors were manually removed (see supplementary material).
Reflecting clinical relevance regarding coarse localisation on medical images and the absence of overlapping objects in 3D images, we report mean Average Precision (mAP) at an IoU threshold of $0.1$ \cite{jaeger_20}.


\noindent\textbf{Data sets.} An overview of all data sets and their properties can be found in the supplementary material. Out of the 13 data sets, we used 10 for development and validation of nnDetection. These are further divided into a training pool (4 data sets: CADA \cite{kossen_20}, LIDC-IDRI \cite{armato_11,jaeger_20}, RibFrac \cite{jin_20} and Kits19 \cite{heller_19}.) and validation pool (6 data sets: ProstateX \cite{litjens_14,cuocolo_21}, ADAM \cite{timmins_20}, Medical Segmentation Decathlon Liver, Pancreas, Hepatic Vessel and Colon \cite{simpson_19}). While in the training pool we used all data for development and report 5-fold cross-validation results, in the validation pool roughly $40\%$ of each data set was split off as held-out test set before development. These test splits were only processed upon final evaluation (for ADAM we used the public leaderboard as hold-out test split). The test pool consists of 3 additional data sets (LUNA16 \cite{setio_2017}, and TCIA Lymph-Node \cite{roth_14,seff_15}) that were entirely held-out from method development to evaluate the generalization ability of our automated method configuration.

\noindent\textbf{Compared methods.} While there exist reference scores in literature for the well-known LUNA16 benchmark and ADAM provides an open leaderboard, there exists no standardized evaluation protocol for object detection methods on the remaining 11 data sets. Thus, we initiate a new benchmark by comparing nnDetection against nnU-Net that we modified to serve as a standardized baseline for object detection (see Section~\ref{sec:nnunet}). This comparison is relevant for three reasons: 1) Segmentation methods are often modified to be deployed on detection tasks in the medical domain \cite{timmins_20}. 2) nnU-Net is currently the only available method that can be readily deployed on a large number of data sets without manual adaptation. 3) The need for tackling medical object detection task with dedicated detection methods rather than segmentation-based substitutes has only been studied on two medical data sets before \cite{jaeger_20}, thus providing large-scale evidence for this comparison is scientifically relevant.

\begin{figure}[t]
    \centering
    \includegraphics[width=\textwidth]{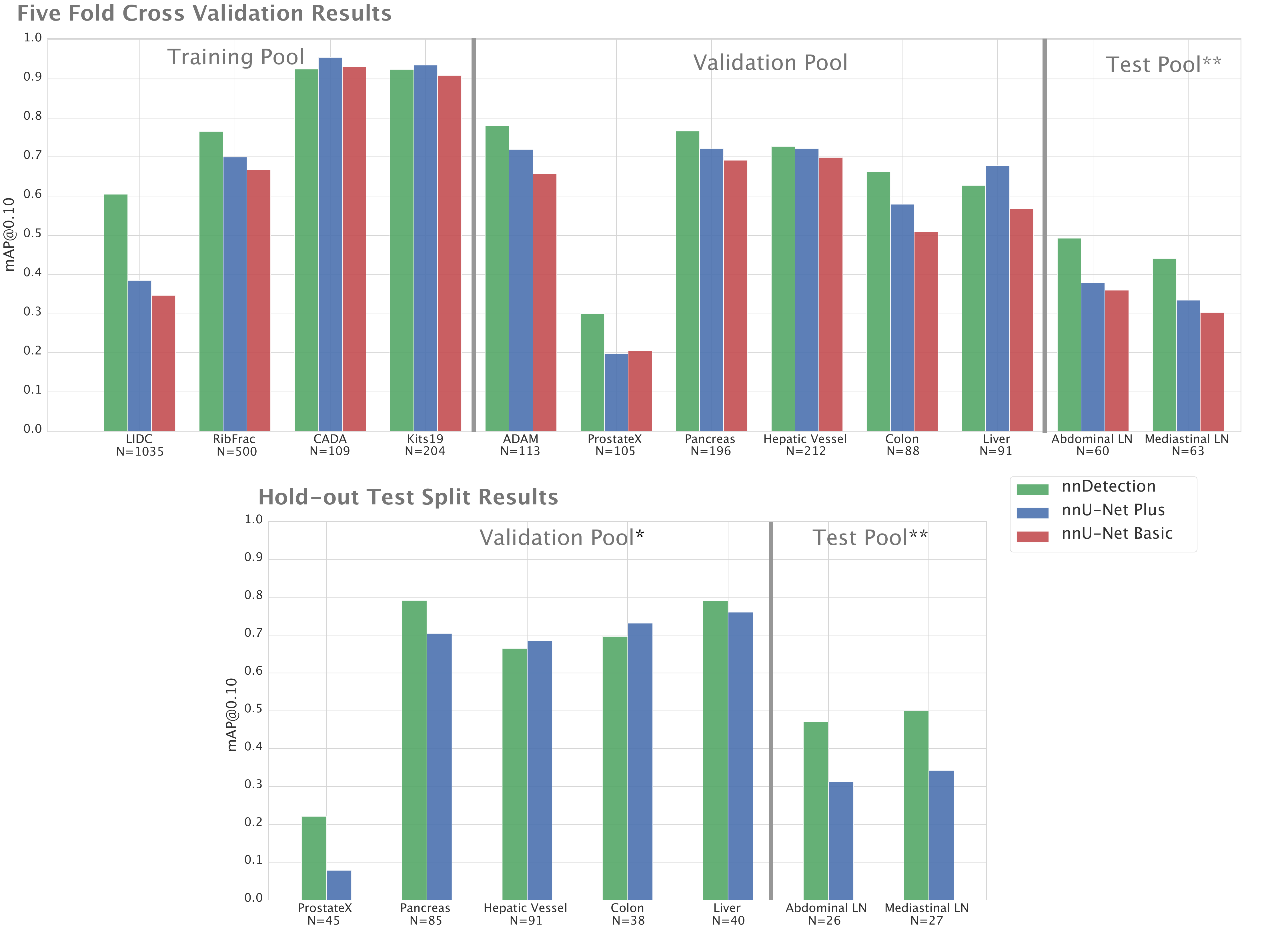}
    \caption{
        Large-scale benchmark against nnU-Net on 12 data sets (cross-validation results on the top and test split results on the bottom panel). $^*$The test split result of ADAM is represented by our submission to the live leaderboard and can be found in the supplementary material). $^{**}$LUNA16 results are visualized in Fig.~\ref{fig:results_luna}. Numerical values for all experiments can be found at \url{https://github.com/MIC-DKFZ/nnDetection}.
        }
    \label{fig:results_pool}
\end{figure}

\noindent\textbf{Public leaderboard results.} LUNA16 \cite{setio_2017} is a long standing benchmark for object detection methods \cite{ding_17,dou_17,khosravan_18,wang_18,zhu_18,liu_19,song_20,cao_20} which consists of 888 CT scans with lung nodule annotations. While LUNA16 images represent a subset of LIDC-IDRI, the task is different since LUNA16 does not differentiate between benign and malignant classes and the annotations differ in that they were reduced to a center point plus radius (for training we generated segmentation labels in the form of circles from this information). As LUNA16 is part of our test pool, nnDetection was applied by simply executing automated method configuration once and without any manual intervention. Our method achieves a Competition Performance Metric (CPM) of $0.930$ outperforming all previous methods on the nodule-candidate-detection task (see Fig.~\ref{fig:results_luna} and supplementary material for details). Our submission to the public leaderboard of the Aneurysm Detection And segMentation (ADAM) \cite{timmins_20} challenge currently ranks third with a sensitivity of $0.64$ at a false positive count of $0.3$ (see supplementary material for more details). One of the two higher ranking submissions is a previous version of nnDetection, which hints upon a natural performance spread on limited test sets in object detection tasks (the previous version represented our original submission to the 2020 MICCAI event, there were only these two submissions to ADAM from our side in total).

\noindent\textbf{Large-scale Comparison against nnU-Net.} nnDetection outperforms the enhanced baseline \textit{nnU-Net Plus} on 9 out of 12 data sets in the cross-validation protocol (Fig~\ref{fig:results_pool} top panel). Thereby, substantial margins ($>5\%$) are observed in 7 data sets and substantially lower performance only on the liver data set \cite{simpson_19}. The baseline with fixed postprocessing strategy (\textit{nnU-Net Basic}) shows worse results than nnDetection on 11 out of 12 data sets. On the hold-out test splits (Fig~\ref{fig:results_pool} bottom panel), nnDetection outperforms \textit{nnU-Net Plus} on 5 out of 7 data sets with substantial margins in 4 of them and only substantially lower performance on the colon data set \cite{simpson_19}. Notably, 4 of the 7 data sets were part of nnU-Net’s development pool and thus not true hold-out splits for the baseline algorithm \cite{isensee_21}. High volatility of evaluation metrics between cross-validation and test results is observed, especially on the liver and colon data sets, hinting upon the importance of evaluation across many diverse data sets.
\section{Discussion}
nnDetection opens a new perspective on method development in medical object detection. All design choices have been optimized on a data set-agnostic meta-level, which allows for out-of-the-box adaptation to specific data sets upon application and removes the burden of manual and iterative method configuration. Despite this generic functionality, nnDetection shows performance superior to or on par with the state-of-the-art on two public leaderboards and 11 benchmarks that were newly proposed for object detection. Our method can be considered as a starting point for further manual task-specific optimization. As seen on LUNA16, an additional false-positive-reduction component can further improve results. Also, data-driven optimization along the lines of AutoML \cite{hutter2019automated} could be computationally feasible for specific components of the object detection pipeline and thus improve results even further.

\noindent In making nnDetection available including models and object annotations for all newly proposed benchmarks we hope to contribute to the rising interest in object detection on medical images by providing a tool for out-of-the-box object predictions, a framework for method development, a standardized baseline to compare against, as well as a benchmark for large-scale method evaluation.
%
%
\section*{Acknowledgements}
Part of this work was funded by the Deutsche Forschungsgemeinschaft (DFG, German Research Foundation) – 410981386 and the Helmholtz Imaging Platform (HIP), a platform of the Helmholtz Incubator on Information and Data Science.

\bibliographystyle{abbrv}
{\small \bibliography{paper1836}}

\begin{thebibliography}{10}

\bibitem{armato_11}
S.~G. Armato~III, G.~McLennan, L.~Bidaut, M.~F. McNitt-Gray, C.~R. Meyer, A.~P.
  Reeves, B.~Zhao, D.~R. Aberle, C.~I. Henschke, E.~A. Hoffman, et~al.
\newblock The lung image database consortium (lidc) and image database resource
  initiative (idri): a completed reference database of lung nodules on ct
  scans.
\newblock {\em Medical physics}, 38(2):915--931, 2011.

\bibitem{cao_20}
H.~{Cao}, H.~{Liu}, E.~{Song}, G.~{Ma}, X.~{Xu}, R.~{Jin}, T.~{Liu}, and C.~C.
  {Hung}.
\newblock A two-stage convolutional neural networks for lung nodule detection.
\newblock {\em IEEE Journal of Biomedical and Health Informatics},
  24(7):2006--2015, 2020.

\bibitem{cuocolo_21}
R.~Cuocolo, A.~Comelli, A.~Stefano, V.~Benfante, N.~Dahiya, A.~Stanzione,
  A.~Castaldo, D.~R.~D. Lucia, A.~Yezzi, and M.~Imbriaco.
\newblock Deep learning whole-gland and zonal prostate segmentation on a public
  mri dataset.
\newblock {\em Journal of Magnetic Resonance Imaging}, 2021.

\bibitem{ding_17}
J.~Ding, A.~Li, Z.~Hu, and L.~Wang.
\newblock Accurate pulmonary nodule detection in computed tomography images
  using deep convolutional neural networks.
\newblock In {\em MICCAI}, pages 559--567. Springer, 2017.

\bibitem{dou_17}
Q.~Dou, H.~Chen, Y.~Jin, H.~Lin, J.~Qin, and P.-A. Heng.
\newblock Automated pulmonary nodule detection via 3d convnets with online
  sample filtering and hybrid-loss residual learning.
\newblock In {\em MICCAI}, pages 630--638. Springer, 2017.

\bibitem{heller_19}
N.~Heller, N.~Sathianathen, A.~Kalapara, E.~Walczak, K.~Moore, H.~Kaluzniak,
  J.~Rosenberg, P.~Blake, Z.~Rengel, M.~Oestreich, et~al.
\newblock The kits19 challenge data: 300 kidney tumor cases with clinical
  context, ct semantic segmentations, and surgical outcomes.
\newblock {\em arXiv preprint arXiv:1904.00445}, 2019.

\bibitem{hutter2019automated}
F.~Hutter, L.~Kotthoff, and J.~Vanschoren.
\newblock {\em Automated machine learning: methods, systems, challenges}.
\newblock Springer Nature, 2019.

\bibitem{isensee_21}
F.~Isensee, P.~F. Jaeger, S.~A. Kohl, J.~Petersen, and K.~H. Maier-Hein.
\newblock nnu-net: a self-configuring method for deep learning-based biomedical
  image segmentation.
\newblock {\em Nature Methods}, 18(2):203--211, 2021.

\bibitem{jaeger_20}
P.~F. Jaeger, S.~A. Kohl, S.~Bickelhaupt, F.~Isensee, T.~A. Kuder, H.-P.
  Schlemmer, and K.~H. Maier-Hein.
\newblock Retina u-net: Embarrassingly simple exploitation of segmentation
  supervision for medical object detection.
\newblock In {\em ML4H}, pages 171--183. PMLR, 2020.

\bibitem{jin_20}
L.~Jin, J.~Yang, K.~Kuang, B.~Ni, Y.~Gao, Y.~Sun, P.~Gao, W.~Ma, M.~Tan,
  H.~Kang, J.~Chen, and M.~Li.
\newblock Deep-learning-assisted detection and segmentation of rib fractures
  from {CT} scans: Development and validation of {FracNet}.
\newblock 62.
\newblock Publisher: Elsevier.

\bibitem{khosravan_18}
N.~Khosravan and U.~Bagci.
\newblock S4nd: Single-shot single-scale lung nodule detection.
\newblock In {\em MICCAI}, pages 794--802. Springer, 2018.

\bibitem{litjens_14}
G.~{Litjens}, O.~{Debats}, J.~{Barentsz}, N.~{Karssemeijer}, and H.~{Huisman}.
\newblock Computer-aided detection of prostate cancer in mri.
\newblock {\em IEEE TMI}, 33(5):1083--1092, 2014.

\bibitem{liu_19}
J.~Liu, L.~Cao, O.~Akin, and Y.~Tian.
\newblock 3dfpn-hs: 3d feature pyramid network based high sensitivity and
  specificity pulmonary nodule detection.
\newblock In {\em MICCAI}, pages 513--521. Springer, 2019.

\bibitem{maier_hein_18}
L.~Maier-Hein, M.~Eisenmann, A.~Reinke, S.~Onogur, M.~Stankovic, P.~Scholz,
  T.~Arbel, H.~Bogunovic, A.~P. Bradley, A.~Carass, C.~Feldmann, A.~F. Frangi,
  P.~M. Full, B.~van Ginneken, A.~Hanbury, K.~Honauer, M.~Kozubek, B.~A.
  Landman, K.~März, O.~Maier, K.~Maier-Hein, B.~H. Menze, H.~Müller, P.~F.
  Neher, W.~Niessen, N.~Rajpoot, G.~C. Sharp, K.~Sirinukunwattana, S.~Speidel,
  C.~Stock, D.~Stoyanov, A.~A. Taha, F.~van~der Sommen, C.-W. Wang, M.-A.
  Weber, G.~Zheng, P.~Jannin, and A.~Kopp-Schneider.
\newblock Why rankings of biomedical image analysis competitions should be
  interpreted with care.
\newblock {\em Nature Communications}, 9(1):5217.
\newblock Number: 1 Publisher: Nature Publishing Group.

\bibitem{redmon_17}
J.~Redmon and A.~Farhadi.
\newblock Yolo9000: better, faster, stronger.
\newblock In {\em CVPR}, pages 7263--7271, 2017.

\bibitem{roth_14}
H.~R. Roth, L.~Lu, A.~Seff, K.~M. Cherry, J.~Hoffman, S.~Wang, J.~Liu,
  E.~Turkbey, and R.~M. Summers.
\newblock A new 2.5 d representation for lymph node detection using random sets
  of deep convolutional neural network observations.
\newblock In {\em MICCAI}, pages 520--527. Springer, 2014.

\bibitem{seff_15}
A.~Seff, L.~Lu, A.~Barbu, H.~Roth, H.-C. Shin, and R.~M. Summers.
\newblock Leveraging mid-level semantic boundary cues for automated lymph node
  detection.
\newblock In {\em MICCAI}, pages 53--61. Springer, 2015.

\bibitem{setio_2017}
A.~A.~A. Setio, A.~Traverso, T.~{de Bel}, M.~S. Berens, C.~van~den Bogaard,
  P.~Cerello, H.~Chen, Q.~Dou, M.~E. Fantacci, B.~Geurts, R.~van~der Gugten,
  P.~A. Heng, B.~Jansen, M.~M. {de Kaste}, V.~Kotov, J.~Y.-H. Lin, J.~T.
  Manders, A.~Sóñora-Mengana, J.~C. García-Naranjo, E.~Papavasileiou,
  M.~Prokop, M.~Saletta, C.~M. Schaefer-Prokop, E.~T. Scholten, L.~Scholten,
  M.~M. Snoeren, E.~L. Torres, J.~Vandemeulebroucke, N.~Walasek, G.~C. Zuidhof,
  B.~van Ginneken, and C.~Jacobs.
\newblock Validation, comparison, and combination of algorithms for automatic
  detection of pulmonary nodules in computed tomography images: The luna16
  challenge.
\newblock {\em MedIA}, 42:1--13, 2017.

\bibitem{simpson_19}
A.~L. Simpson, M.~Antonelli, S.~Bakas, M.~Bilello, K.~Farahani,
  B.~Van~Ginneken, A.~Kopp-Schneider, B.~A. Landman, G.~Litjens, B.~Menze,
  et~al.
\newblock A large annotated medical image dataset for the development and
  evaluation of segmentation algorithms.
\newblock {\em arXiv preprint arXiv:1902.09063}, 2019.

\bibitem{song_20}
T.~Song, J.~Chen, X.~Luo, Y.~Huang, X.~Liu, N.~Huang, Y.~Chen, Z.~Ye, H.~Sheng,
  S.~Zhang, and G.~Wang.
\newblock {CPM}-net: A 3d center-points matching network for pulmonary nodule
  detection in {CT} scans.
\newblock In A.~L. Martel, P.~Abolmaesumi, D.~Stoyanov, D.~Mateus, M.~A.
  Zuluaga, S.~K. Zhou, D.~Racoceanu, and L.~Joskowicz, editors, {\em MICCAI},
  pages 550--559. Springer International Publishing.

\bibitem{kossen_20}
C.~Tabea~Kossen, L.~Kaufhold, M.~Hüllebrand, J.-M. Kuhnigk, J.~Brühning,
  J.~Schaller, B.~Pfahringer, A.~Spuler, L.~Goubergrits, and A.~Hennemuth.
\newblock Cerebral aneurysm detection and analysis, Mar. 2020.

\bibitem{timmins_20}
K.~Timmins, E.~Bennink, I.~van~der Schaaf, B.~Velthuis, Y.~Ruigrok, and
  H.~Kuijf.
\newblock {Intracranial Aneurysm Detection and Segmentation Challenge}, Mar.
  2020.

\bibitem{wang_18}
B.~Wang, G.~Qi, S.~Tang, L.~Zhang, L.~Deng, and Y.~Zhang.
\newblock Automated pulmonary nodule detection: High sensitivity with few
  candidates.
\newblock In {\em MICCAI}, pages 759--767. Springer, 2018.

\bibitem{zhang_20}
S.~Zhang, C.~Chi, Y.~Yao, Z.~Lei, and S.~Z. Li.
\newblock Bridging the gap between anchor-based and anchor-free detection via
  adaptive training sample selection.
\newblock In {\em CVPR}, pages 9759--9768, 2020.

\bibitem{zhu_18}
W.~Zhu, C.~Liu, W.~Fan, and X.~Xie.
\newblock Deeplung: Deep 3d dual path nets for automated pulmonary nodule
  detection and classification.
\newblock In {\em WACV}, pages 673--681. IEEE, 2018.

\bibitem{zlocha_19}
M.~Zlocha, Q.~Dou, and B.~Glocker.
\newblock Improving retinanet for ct lesion detection with dense masks from
  weak recist labels.
\newblock In {\em MICCAI}, pages 402--410. Springer, 2019.

\end{thebibliography}

\newpage
\appendix
\section{Supplementary Material}

\begin{figure}[htb]
    \centering
    \includegraphics[width=0.75\linewidth]{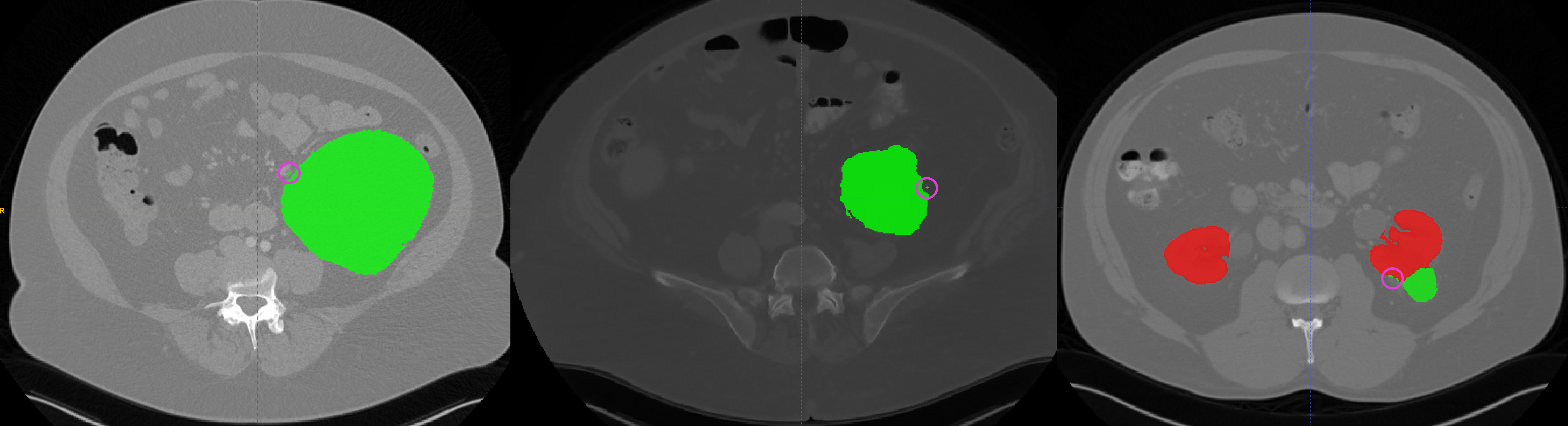}
    \caption{
    Shows examples (voxels inside pink circles) of small clusters (typically only visible in one slice) which were removed from the annotations and not considered as individual objects from Kits19\cite{heller_19}.
    Red annotations mark the kidney and were not used for training (neither nnDetection nor nnU-Net) while the green annotations denote tumor regions. Note, that this procedure was not performed for the Decathlon Liver data set which contained too many small objects.
    }
    \label{fig:manual_removal}
\end{figure}

\begin{figure}[htb]
    \centering
    \includegraphics[width=0.75\linewidth]{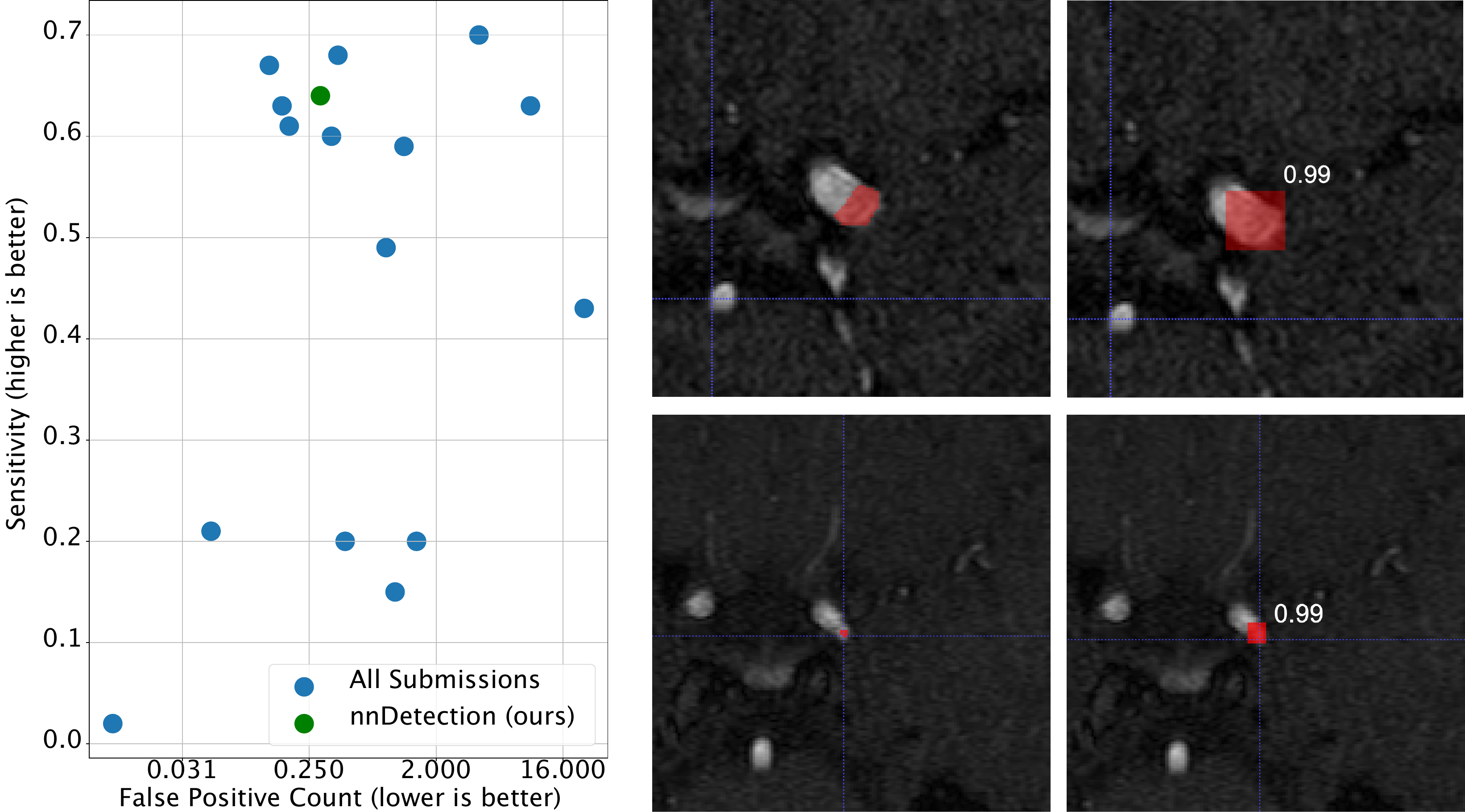}
    \caption{
    \textit{Left:} Overview ADAM challenge live leaderboard.
    nnDetection currently ranks third with a sensitivity of $0.64$ at an false positive count of $0.3$ (top left corner is the best).
    \textit{Right:} The left images show examples for the ground truth aneurysm annotation.
    The right images show the predictions of nnDetection.
    Especially small aneurysms have a very small tolerance for predictions to be regarded as true positives on the live leaderboard due to their very small radius.
    }
    \label{fig:results_adam}
\end{figure}

\begin{table}
\centering
    \begin{tabular}{ | c | c | c | c | c | c | c | } 
    \hline
        Name
        & Pool
        & \multicolumn{1}{|p{2cm}|}{\centering Original \\ Object \\ Labels} 
        & \multicolumn{1}{|p{1.5cm}|}{\centering Manually \\ Checked}
        & \multicolumn{1}{|p{1.5cm}|}{\centering Number \\ of Scans \\ (Tr/Ts)}
        & \multicolumn{1}{|p{1.5cm}|}{\centering FG \\ Classes}
        & Split \\
    \hline
        CADA
        & Train
        & Yes
        & NA
        & 109
        & aneurysm
        & Random \\
    \hline
        LIDC
        & Train
        & Yes
        & NA
        & 1035
        & \begin{tabular}[c]{@{}c@{}}benign lesion, \\ mal. lesion\end{tabular} 
        & Custom \\
    \hline
        RibFrac
        & Train
        & Yes
        & NA
        & 500
        & rib fractures
        & Random \\
    \hline
        Kits19
        & Train
        & No
        & Yes
        & 204
        & tumor
        & Random \\
    \hline
    \hline
        ADAM
        & Val
        & Yes
        & NA
        & 113
        & aneurysm
        & Patient \\
    \hline
        ProstateX
        & Val
        & Yes
        & NA
        & 140 / 60
        & \begin{tabular}[c]{@{}c@{}}significant \\ lesion, \\ insig. lesion\end{tabular} 
        & Random \\
    \hline
        Liver
        & Val
        & No
        & No
        & 91 / 40
        & tumor
        & Random \\
    \hline
        Pancreas
        & Val
        & No
        & Yes
        & 196 / 85
        & tumor
        & Random \\
    \hline
        Hepatic Vessel
        & Val
        & No
        & Yes
        & 212 / 91
        & tumor
        & Random \\
    \hline
        Colon
        & Val
        & No
        & Yes
        & 88 / 38
        & tumor
        & Random \\
    
    \hline
    \hline
        \begin{tabular}[c]{@{}c@{}} Abdominal  \\ Lymph Nodes\end{tabular}
        & Test
        & Yes
        & NA
        & 60 / 26
        & \begin{tabular}[c]{@{}l@{}}lymph  \\ nodes\end{tabular}
        & Random \\
    \hline
        \begin{tabular}[c]{@{}c@{}} Mediastinal  \\ Lymph Nodes\end{tabular}
        & Test
        & Yes
        & NA
        & 63 / 27
        & \begin{tabular}[c]{@{}l@{}}lymph  \\ nodes\end{tabular}
        & Random \\
    \hline
        LUNA
        & Test
        & Yes
        & NA
        & 888
        & nodules
        & Official \\
    \hline
    \end{tabular}
    \caption{
    Pool and dataset overview.
    For datasets which did not have object labels, we applied connected components to the semantic segmentation annotations and discarded all objects with a diameter less than $3mm$ followed by a manual check to remove obvious mistakes (see. Fig~\ref{fig:manual_removal}). NA was inserted if original object labels were available.
    For datasets which had additional organ annotations only the tumor label was used for training nnDetection and nnU-Net.
    We used a patient stratified split for ADAM and a custom split for LIDC \cite{jaeger_20}.
    }
    \label{tab:pool_overview}
\end{table}

\begin{table}
    \centering
    \begin{tabular}{ | c | c | m{2.7em} | m{2.7em} | m{2.7em} | m{2.7em} | m{2.7em} | m{2.7em} | m{2.7em} | m{2.57em} |} 
        \hline
        Methods & 1/8 & 1/4 & 1/2 & 1 & 2 & 4 & 8 & CPM \\
        \hline
        Dou et al.   (2017a)    & 0.659 & 0.745 & 0.819 & 0.865 & 0.906 & 0.933 & 0.946 & 0.839 \\
        Zhu et al.   (2018)     & 0.692 & 0.769 & 0.824 & 0.865 & 0.893 & 0.917 & 0.933 & 0.842 \\
        Wang et al.  (2018)     & 0.676 & 0.776 & 0.879 & 0.949 & 0.958 & 0.958 & 0.958 & 0.878 \\
        Ding et al.  (2017)     & 0.748 & 0.853 & 0.887 & 0.922 & 0.938 & 0.944 & 0.946 & 0.891 \\
        Khosravan et al. (2018) & 0.709 & 0.836 & 0.921 & \textbf{0.953} & 0.953 & 0.953 & 0.953 & 0.897 \\
        Liu et al. (2019)       & \textbf{0.848} & 0.876 & 0.905 & 0.933 & 0.943 & 0.957 & 0.970 & 0.919 \\
        Song et al. (2020)      & - & - & - & 0.911 & 0.928 & - & 0.948 & - \\
        \hline
        nnDetection(ours)       & 0.812 & \textbf{0.885} & \textbf{0.927} & 0.950 & \textbf{0.969} & \textbf{0.979} & \textbf{0.985} & \textbf{0.930} \\
        \hline
        \hline
        Cao et al. (2020) + FPR   & 0.848 & 0.899 & 0.925 & 0.936 & 0.949 & 0.957 & 0.960 & 0.925 \\
        Liu et al. (2019) + FPR   & 0.904 & 0.914 & 0.933 & 0.957 & 0.971 & 0.971 & 0.971 & 0.952 \\
        \hline
    \end{tabular}
    \caption{Shows the sensitivity values at the predefined false positive per scan thresholds of the LUNA16 challenge.
    nnDetection outperforms all methods which do not employ an additional False Positive Reduction stage.
    Only Liu et al. \cite{liu_19} \textbf{with} FPR achieve an higher CPM score, especially due to improved performance at low
    false positive per scan thresholds.}
    \label{tab:results_luna}
\end{table}

\end{document}